\documentclass[10pt,twocolumn]{IEEEtran}

\usepackage{graphicx}
\usepackage{amsmath}
\usepackage{amssymb}
\usepackage[caption=false]{subfig}
\usepackage[noadjust]{cite}
\usepackage{float}
\usepackage{algorithm}
\usepackage{bibentry}
\usepackage{balance}
\usepackage{algorithm}
\usepackage{algorithmicx}
\usepackage{algpseudocode}
\usepackage{xcolor}

\graphicspath{{img/}}

\begin{document}

\bstctlcite{IEEEexample:BSTcontrol}

\title{Hybrid Precoding for mmWave Massive MIMO with One-Bit DAC}

\author{Sung Joon Maeng, Yavuz Yap{\i}c{\i}, \.{I}smail G\"{u}ven\c{c}, Huaiyu Dai, and Arupjyoti Bhuyan}%

\maketitle

\begin{abstract}
Hybrid beamforming is key to achieving energy-efficient 5G wireless networks equipped with massive amount of antennas. Low-resolution data converters bring yet another degree of freedom to energy efficiency for the state-of-the-art 5G transceivers. In this work, we consider the design of hybrid precoders for massive multiple-input multiple-output (MIMO) channels in millimeter-wave (mmWave) spectrum along with one-bit digital-to-analog converters (DACs) and finite-quantized phase shifters. In particular, we propose an alternating-optimization-based precoder design which recursively computes the covariance of the quantization distortion, and updates the precoders accordingly. Numerical results verify that the achievable rate improves quickly through iterations that involve updates to the weight matrix, distortion covariance of the quantization, and the respective precoders. 
\end{abstract}


\vspace{-0.12in}

\section{Introduction}
The vastly unoccupied mmWave frequency band has been envisioned as a promising solution to spectrum scarcity over the conventional sub-6GHz communications \cite{Rappaport2017OveMil}. The high path loss in mmWave frequencies should be compensated by effective beamforming strategies. Thanks to the form factors getting smaller in mmWave spectrum, it becomes possible to squeeze large antenna arrays even in mobile devices, and obtain sufficiently large beamforming gains. The energy efficiency, however, emerges as a concern when employing large antenna arrays, since conventional  implementation (i.e, full-digital) requires a dedicated data converter (i.e., analog-to-digital converter (ADC), DAC) and radio-frequency (RF) chain (e.g., mixer, oscillator) for each antenna element. The hybrid beamforming, on the other hand, splits the overall precoding into the baseband and RF stages, which in turn cuts down the required number of data converters and RF chains, and, hence, improves energy efficiency \cite{Heath2014SpaSpa}.

The data converters with large bandwidth than ever before, which are particularly critical in mmWave communications, require exponentially increasing power consumption for integer number of resolution bits \cite{Swindlehurst2017AnaOne}. At the transmitter side with relatively larger antenna arrays, it is even more challenging to achieve desired beamforming gains at moderate power budgets. In this work, we therefore consider the hybrid precoder design for mmWave transmitters assuming one-bit DAC and massive MIMO. 

There are limited number of works in the literature considering hybrid precoding along with low-resolution data converters, which generally assume low-resolution data converters at the receiver side (i.e., few-bit ADCs, infinite-resolution DACs). In particular, \cite{Heath2017HybArc} considers two different hybrid precoder designs, which are based on channel inversion and singular value decomposition (SVD), for mmWave channels with few-bit ADCs. A hybrid precoder is proposed in \cite{Dongliang2018HybPre} for point-to-point MIMO downlink assuming spatially uncorrelated channel and one-bit ADCs. In a recent study of \cite{Rupp2018EneEff}, a hybrid precoder design based solely on SVD is proposed for spatially uncorrelated MIMO channels with few-bit DACs.

In this work, we propose a novel hybrid precoder for spatially correlated mmWave channels with one-bit DAC and finite-quantized phase shifters. In particular, contribution of the quantization distortion is carefully taken into account through recursive computation of the respective covariance matrix (instead of assuming a zero covariance \cite{Rupp2018EneEff}), which produces more reliable achievable rates. As a complete precoder design, we develop a novel alternating-maximization strategy where the baseband and RF precoders are updated iteratively relying on the Bussgang theorem \cite{Bussgang1952Cross}, which is superior to additive quantization noise model (AQNM) \cite{Erkip2015LowPow} adopted by  \cite{Heath2017HybArc,Dongliang2018HybPre,Rupp2018EneEff}. The numerical results verify the effectiveness of the proposed precoder design.
 
\vspace{-0.1in}

\section{System Model} 
\label{sec:system}

\begin{figure}[!t]
	\centering
	\vspace{-0.0in}
	\includegraphics[width=0.47\textwidth]{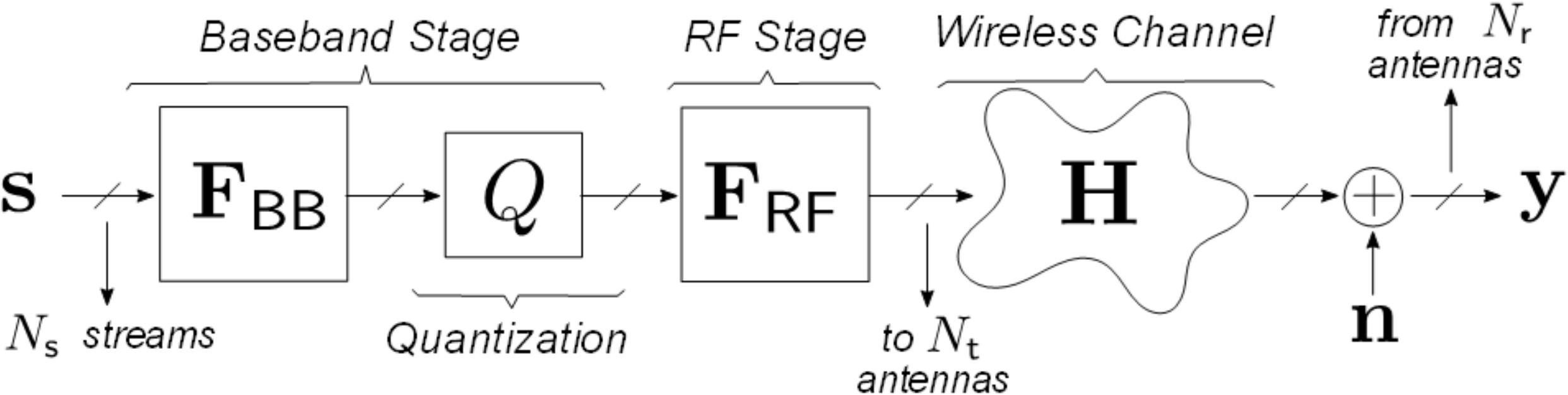}
	\vspace{-0.1in}
	\caption{System model for the point-to-point mmWave MIMO downlink with hybrid precoding and one-bit DAC.}
	\label{fig:setting}
	\vspace{-0.2in}
\end{figure}

We consider a point-to-point massive MIMO downlink in mmWave spectrum, as illustrated in Fig.~\ref{fig:setting}, where a transmitter equipped with $N_\mathsf{t}$ antennas and $N_\mathsf{RF}$ RF chains sends $N_\mathsf{s}$ data steams to a receiver with $N_\mathsf{r}$ antennas. Considering $N_\mathsf{t}$ being typically large, we assume a hybrid precoder at the transmitter with one-bit DACs to relieve the complexity and improve the energy efficiency, and infinite-resolution ADCs at the receiver, thanks to relatively smaller $N_\mathsf{r}$. 

Assuming $\textbf{F}_\mathsf{RF} \,{\in}\, \mathbb{C}^{N_\mathsf{t}{\times}N_\mathsf{RF}}$, $\textbf{F}_\mathsf{BB} \,{\in}\, \mathbb{C}^{N_\mathsf{RF}{\times} N_\mathsf{s}}$, and $\textbf{s} \,{\in}\, \mathbb{C}^{N_\mathsf{s}\times 1}$ are the analog RF precoder, digital baseband precoder, and transmitted data, respectively, the received signal is 
\begin{align}\label{eq:observation}
\textbf{y} &= \textbf{H}\hspace{0.01in}\textbf{F}_\mathsf{RF}\mathcal{Q}(\textbf{F}_\mathsf{BB}\textbf{s}) + \textbf{n},
\end{align}
where $\mathcal{Q}(\cdot)$ is one-bit quantization, and $\textbf{n}$ is the observation noise composed of independent complex Gaussian entries with zero-mean and variance $\sigma^2_\mathsf{n}$. We also assume that $\mathbb{E}[\textbf{s}\textbf{s}^{\rm H}] \,{=}\, \frac{P_s}{N_\mathsf{s}}\textbf{I}_{N_\mathsf{s}}$ where $P_\mathsf{s}$ is the total power of the unprecoded data, and $\textbf{I}_{N_\mathsf{s}} \,{\in}\, \mathbb{R}^{N_\mathsf{s}{\times}N_\mathsf{s}}$ is the identity matrix. In \eqref{eq:observation}, $\textbf{H} \,{\in}\, \mathbb{C}^{N_\mathsf{r}{\times} N_\mathsf{t}}$ represents the spatially correlated mmWave channel with $N_\mathsf{c}$ clusters and $N_\mathsf{p}$ rays \cite{Rappaport2017OveMil}.


\section{Linear Quantization Models} \label{sec:linearization}

We consider two linearization schemes for the non-linear quantization operator $\mathcal{Q}(\cdot)$ while deriving the desired precoders. The first scheme is AQNM \cite{Erkip2015LowPow}, which is widely adopted---thanks to its simplicity---although it is known to overestimate the achievable rate performance especially for the small number of quantization bits
\cite{Heath2017HybArc,Zhang2017OnTra}. The respective linearization is expressed as follows
\begin{align}\label{eq:AQNM}
\mathcal{Q}(\textbf{F}_\mathsf{BB}\textbf{s}) & \approx \textbf{A}_\mathsf{Q}\textbf{F}_\mathsf{BB}\textbf{s} + \textbf{q}_\mathsf{Q},
\end{align}
where $\textbf{A}_\mathsf{Q}$ is the weight matrix given as
\begin{align}\label{eq:A_AQNM}
    \textbf{A}_\mathsf{Q} &= \sqrt{1-\eta_b}\textbf{I}_{N_\mathsf{s}},
\end{align}
with $\eta_b$ being the distortion factor, which is generally approximated by $\eta_b \,{\approx}\, \frac{\pi\sqrt{3}}{2}2^{{-}2b}$ for $b$-bit quantization with sufficiently large $b$, while its more accurate value for one-bit quantization is $\eta_b \,{\approx}\, 0.3634$ \cite{Max1960QuaMin}. In \eqref{eq:AQNM}, $\textbf{q}_\mathsf{Q}$ stands for the quantization distortion with the covariance 
\begin{align}\label{eq:Cqq_AQNM}
\textbf{C}_{\textbf{q}_\mathsf{Q}\textbf{q}_\mathsf{Q}}&=\mathbb{E}[\textbf{q}_\mathsf{Q}\textbf{q}_\mathsf{Q}^{\rm H}]=\frac{P_s}{N_\mathsf{s}}\eta_b(1-\eta_b)\text{diag}\left(\textbf{F}_\mathsf{BB}\textbf{F}_\mathsf{BB}^{\rm H}\right),
\end{align}
which seemingly ignores the correlation between the entries of $\textbf{q}_\mathsf{Q}$ (i.e., $\textbf{C}_{\textbf{q}_\mathsf{Q}\textbf{q}_\mathsf{Q}}$ is diagonal) that is the basic reason behind its not-sufficiently-accurate rates.

The second model we consider is based on the Bussgang theorem \cite{Bussgang1952Cross}, which suggest the following decomposition
\begin{align}\label{eq:Bussgang}
\mathcal{Q}(\textbf{F}_\mathsf{BB}\textbf{s})\approx\textbf{A}_\mathsf{B}\textbf{F}_\mathsf{BB}\textbf{s} + \textbf{q}_\mathsf{B},
\end{align}
where $\textbf{A}_\mathsf{B}$ is the weight matrix given as
\begin{align}\label{eq:A_Bussgang}
\textbf{A}_\mathsf{B} \,{=} \sqrt{\frac{2}{\pi }}\left[\text{diag}\left(\textbf{C}_{\textbf{x}\textbf{x}}\right)\right]^{{-}\frac{1}{2}} {=} \sqrt{\frac{2N_\mathsf{s}}{\pi P_s}}\left[\text{diag}\left(\textbf{F}_\mathsf{BB}\textbf{F}_\mathsf{BB}^{\rm H}\right)\right]^{{-}\frac{1}{2}},
\end{align}
with $\textbf{C}_{\textbf{x}\textbf{x}}$ being the covariance of $\textbf{F}_\mathsf{BB} \textbf{s}$. In \eqref{eq:Bussgang}, $\textbf{q}_\mathsf{B}$ is the quantization noise with the covariance
\begin{align}\label{eq:Cqq_Bussgang}
\textbf{C}_{\textbf{q}_\mathsf{B}\textbf{q}_\mathsf{B}}&=\textbf{C}_{\textbf{x}_\mathsf{q}\textbf{x}_\mathsf{q}}-\textbf{A}_\mathsf{B}\textbf{C}_{\textbf{x}\textbf{x}}\textbf{A}_\mathsf{B},
\end{align}
where $\textbf{C}_{\textbf{x}_\mathsf{q}\textbf{x}_\mathsf{q}}$ is obtained through arcsin law \cite[eq.~(14)]{Yapici2019QuanMassC}. Note that \eqref{eq:Cqq_Bussgang} does not ignore any correlation in $\textbf{q}_\mathsf{B}$, and hence achieves a superior performance as compared to \eqref{eq:Cqq_AQNM}.  

\section{Hybrid Beamforming for One-Bit DACs} \label{sec4}

In order to derive the desired precoders $\textbf{F}_\mathsf{RF}$ and $\textbf{F}_\mathsf{BB}$, we first incorporate the linear models \eqref{eq:AQNM} and \eqref{eq:Bussgang} into the observation model \eqref{eq:observation}, which can be jointly represented as
\begin{align}\label{eq:observation_modified}
\textbf{y} &= \textbf{H}_\mathsf{e}  \textbf{F}_\mathsf{BB}\textbf{s} + \Tilde{\textbf{n}}, 
\end{align}
where $\textbf{H}_\mathsf{e} \,{=}\, \textbf{H}\hspace{0.01in}\textbf{F}_\mathsf{RF} \textbf{A}_\mathsf{c}$ is the effective channel seen by the baseband precoder, and $\Tilde{\textbf{n}} \,{=}\, \textbf{H}\textbf{F}_\mathsf{RF}\textbf{q}_\mathsf{c} \,{+}\, \textbf{n}$ is the aggregate noise composed of observation noise and scaled quantization noise, with the subscript $\mathsf{c} \,{\in}\, \{\mathsf{Q},\mathsf{B}\}$. In the rest of the paper, we drop the subscript of $\textbf{A}_\mathsf{c}$ and $\textbf{C}_{\textbf{q}_\mathsf{c}\textbf{q}_\mathsf{c}}$ whenever the expressions are common to AQNM and the Bussgang schemes.

Although $\Tilde{\textbf{n}}$ is not necessarily Gaussian, a lower bound on the achievable rate---considering the fact that the mutual information is the worst for Gaussian noise \cite{Nossek2012CapLow}---is given as
\begin{align}\label{eq:rate}
\mathsf{R_{ach}} &= \log_2 \left|\textbf{I}_{N_\mathsf{r}}+\frac{P_s}{N_\mathsf{s}}\textbf{C}_{\Tilde{\textbf{n}}\Tilde{\textbf{n}}}^{-1}\textbf{H}_\mathsf{e}\textbf{F}_\mathsf{BB}\textbf{F}_\mathsf{BB}^{\rm H}\textbf{H}_\mathsf{e}^{\rm H}\right|,
\end{align}
where $\textbf{C}_{\Tilde{\textbf{n}}\Tilde{\textbf{n}}}$ is the covariance of $\Tilde{\textbf{n}}$ given by
\begin{align} \label{eq:covariance_aggregate_noise}
\textbf{C}_{\Tilde{\textbf{n}}\Tilde{\textbf{n}}}&=\textbf{H}\textbf{F}_\mathsf{RF}\textbf{C}_{\textbf{q}\textbf{q}}\textbf{F}_\mathsf{RF}^{\rm H}\textbf{H}^{\rm H}+\sigma^2_\mathsf{n}\textbf{I}_{N_\mathsf{r}}.
\end{align}
The optimization problem to obtain $\textbf{F}_\mathsf{RF}$ and $\textbf{F}_\mathsf{BB}$ can then be formulated to maximize the achievable rate in \eqref{eq:rate} as follows
\begin{IEEEeqnarray}{rl}
\max_{\textbf{F}_\mathsf{RF}, \textbf{F}_\mathsf{BB}}
&\quad \mathsf{R_{ach}} , \label{eq:global_optimization} \\
\text{s.t.}
&\quad \|\textbf{F}_\mathsf{RF} \mathcal{Q} \left( \textbf{F}_\mathsf{BB} \textbf{s} \right) \|^2_\mathsf{F} \leq \mathsf{P}_\mathsf{max},  \IEEEyessubnumber \label{eq:global_optimization_1}\\
&\quad |[\textbf{F}_\mathsf{RF}]_{m,n}|=\frac{1}{\sqrt{N_\mathsf{t}}}, \forall \, m,n,
\IEEEyessubnumber \label{eq:global_optimization_2}
\end{IEEEeqnarray}
where \eqref{eq:global_optimization_1} is the power constraint taking into account \eqref{eq:observation} and the maximum transmit signal power $\mathsf{P}_\mathsf{max}$, and \eqref{eq:global_optimization_2} is due to the assumption that the RF precoder is composed of finite-quantized phase shifters. Incorporating \eqref{eq:AQNM} and \eqref{eq:Bussgang}, and the fact that data and quantization noise is independent of the transmitted data, the power of the transmitted data becomes
\begin{align}
    \!\!\!\|\textbf{F}_\mathsf{RF} \mathcal{Q} \left( \textbf{F}_\mathsf{BB} \textbf{s} \right)\|^2_\mathsf{F} &\,{=}\, \frac{P_s}{N_\mathsf{s}}\|\textbf{F}_\mathsf{RF}\textbf{A}\textbf{F}_\mathsf{BB}\|^2_\mathsf{F}+{\rm tr} \left( \textbf{F}_\mathsf{RF}\textbf{C}_{\textbf{q}\textbf{q}}\textbf{F}_\mathsf{RF}^{\rm H} \right) \!, \label{eq:power_constraint_1}\\
    &\,{=}\, \frac{P_s}{N_\mathsf{s}}\|\textbf{A}\textbf{F}_\mathsf{BB}\|^2_\mathsf{F}+{\rm tr} \left(\textbf{C}_{\textbf{q}\textbf{q}}\right), \label{eq:power_constraint_2}
\end{align}
which follows from the assumption of semi-unitary $\textbf{F}_\mathsf{RF}$ \cite{Heath2017HybArc}.

\section{Iterative Precoder Design}
We adopt alternating-optimization strategy to optimize $\textbf{F}_\mathsf{RF}$ and $\textbf{F}_\mathsf{BB}$ separately, rather than solving \eqref{eq:global_optimization} for these two precoders jointly which is computationally far too expensive. 

\vspace{-0.15in}

\subsection{RF Precoder Design} \label{sec:rf_precoder}

We compute the initial $\textbf{F}_\mathsf{RF}$ using SVD of the channel matrix, i.e., $\textbf{H} \,{=}\, \textbf{U} \boldsymbol{\Sigma} \textbf{V}^{\rm H}$. In particular, we form the candidate RF precoder $\hat{\textbf{F}}_\mathsf{RF}$ by using $N_\mathsf{RF}$ columns of $\textbf{V}$ which correspond to the largest $N_\mathsf{RF}$ singular values of $\textbf{H}$ (i.e., diagonals of $\boldsymbol{\Sigma}$). Although columns of $\textbf{V}$ satisfy constant-norm constraint in \eqref{eq:global_optimization_2}, we employ the well-known alternating projection algorithm (APA) to meet semi-unitary constraint \cite{Heath2017HybArc}, as well. This algorithm iteratively projects the RF precoder onto the vector space composed of constant-norm matrices, and projects it back to the semi-unitary matrix space until a convergence criterion is met.

The SVD-based RF precoder design is suitable for the initial iteration of the alternating-optimization precoder design where the baseband precoder $\textbf{F}_\mathsf{BB}$, covariance of the quantization distortion $\textbf{C}_{\textbf{q}_\mathsf{B}\textbf{q}_\mathsf{B}}$, and weight matrix $\textbf{A}_\mathsf{B}$ are all yet to be computed. Once $\textbf{F}_\mathsf{BB}$ is obtained in the baseband precoder design phase, $\textbf{C}_{\textbf{q}_\mathsf{B}\textbf{q}_\mathsf{B}}$ and $\textbf{A}_\mathsf{B}$ can be computed by \eqref{eq:A_Bussgang} and \eqref{eq:Cqq_Bussgang}, respectively. Then, we propose a new RF precoder that is redesigned so as to maximize the achievable rate directly (instead of relying on SVD of the channel as in the initial iteration) as follows
\begin{IEEEeqnarray}{rl}
\max_{\varphi_{mn} \in \,\mathcal{S}_\varphi \, \forall \, m,n}
&\quad \mathsf{R_{ach}} ,  \label{eq:rf_optimization} \\
\text{s.t.}
&\quad \frac{P_s}{N_\mathsf{s}}\|\textbf{A}_\mathsf{B}\textbf{F}_\mathsf{BB}\|^2_\mathsf{F}\,{+}\,{\rm tr} \left(\textbf{C}_{\textbf{q}_\mathsf{B}\textbf{q}_\mathsf{B}}\right) \leq \mathsf{P}_\mathsf{max}, \IEEEeqnarraynumspace \IEEEyessubnumber \label{eq:rf_optimization_1}
\end{IEEEeqnarray}
where $\varphi_{mn}$ is the finite-quantized phase of the $(m,n)$th entry of the RF precoder (i.e., $[\textbf{F}_\mathsf{RF}]_{m,n} \,{=}\,\frac{1}{N_\mathsf{t}}{\rm e}^{j\varphi_{mn}}$), and $\mathcal{S}_\varphi \,{=}\, \{-\pi,-\pi+\Delta,\dots,\pi\}$ with $\Delta$ being the resolution of the phase shifters. Once the candidate RF precoder is found by \eqref{eq:rf_optimization}, we carry out APA to satisfy the semi-unitary condition.

Note that although there are various methods to solve the optimization problem in \eqref{eq:rf_optimization} for the optimal RF precoder, we adopt a greedy based approach seeking for the best phase values for each entry separately \cite{Dongliang2018HybPre}. The respective number of searches is $\lceil\frac{2\pi}{\Delta}\rceil {\times} N_\mathsf{t} {\times} N_\mathsf{RF}$, which is not prohibitive for moderate array size and $\Delta$. Note also that we employ the weight matrix $\textbf{A}_\mathsf{B}$ and the covariance $\textbf{C}_{\textbf{q}_\mathsf{B}\textbf{q}_\mathsf{B}}$, which are both based on the Bussgang theorem, in \eqref{eq:rf_optimization} after the initial iteration of the alternating-optimization precoder design. By this way, the proposed design procedure gets rid of the shortcoming of AQNM, which ignores the correlation between entries of the quantization distortion, as discussed in Section~\ref{sec:linearization}. 

\vspace{-0.1in}

\subsection{Baseband Precoder Design} \label{sec:bb_precoder}

We adopt an SVD-based strategy to find the optimal baseband precoder $\textbf{F}_\mathsf{BB}$. Different from the initial SVD-based RF precoder design, the effective channel seen by the baseband precoder is now represented by $\textbf{H}_\mathsf{e}$ in \eqref{eq:observation_modified}, which is composed of not only the channel $\textbf{H}$ but also the RF precoder $\textbf{F}_\mathsf{RF}$ and the weight matrix $\textbf{A}_\mathsf{c}$, with $\mathsf{c} \,{\in}\, \{\mathsf{Q,B}\}$. We therefore take into account $\textbf{H}_\mathsf{e}$ while designing the baseband precoder, which is different from \cite{Rupp2018EneEff} where any contribution of $\textbf{F}_\mathsf{RF}$ and $\textbf{A}_\mathsf{c}$ are ignored. In particular, we compute $\textbf{H}_\mathsf{e} \,{=}\, \textbf{U}_\mathsf{e} \boldsymbol{\Sigma}_\mathsf{e} \textbf{V}_\mathsf{e}^{\rm H}$, and use the first $N_\mathsf{s}$ columns of $\textbf{V}_\mathsf{e}$ to obtain a candidate baseband precoder
$\hat{\textbf{F}}_\mathsf{BB}$
assuming the diagonals $\boldsymbol{\Sigma}_\mathsf{e}$ are sorted in descending order. 

Note that since $\textbf{A}_\mathsf{B}$ of the Bussgang theorem given in \eqref{eq:A_Bussgang} is a direct function of $\textbf{F}_\mathsf{BB}$, which is yet to be designed for the initial iteration, we employ $\textbf{A}_\mathsf{Q}$ of AQNM given in \eqref{eq:A_AQNM} while obtaining $\textbf{H}_\mathsf{e}$ since it is readily available. Once $\textbf{F}_\mathsf{BB}$ becomes available from the previous iterations, we use $\textbf{A}_\mathsf{B}$ in $\textbf{H}_\mathsf{e}$ to seek a better performance. Note also that the final baseband precoder $\textbf{F}_\mathsf{BB}$ should satisfy the power constraint in \eqref{eq:power_constraint_2}. We therefore normalize the power of  candidate precoder $\hat{\textbf{F}}_\mathsf{BB}$ as 
\begin{align}\label{eq:bb_normalization}
    \textbf{F}_\mathsf{BB} = \left[{\frac{\mathsf{P}_\mathsf{max}-{\rm tr} \left(\textbf{C}_{\textbf{q}_\mathsf{Q}\textbf{q}_\mathsf{Q}}\right)} {\frac{P_s}{N_\mathsf{s}}\|\textbf{A}_\mathsf{c}\hat{\textbf{F}}_\mathsf{BB}\|^2_\mathsf{F}}}\right]^{\frac12} \hat{\textbf{F}}_\mathsf{BB},
\end{align}
where $\mathsf{c} \,{=}\, \mathsf{Q}$ ($\mathsf{c} \,{=}\, \mathsf{B}$) for the first (subsequent) iteration(s).

As discussed in Section~\ref{sec:linearization}, the covariance in \eqref{eq:Cqq_Bussgang} is superior to \eqref{eq:Cqq_AQNM} as it takes into account the correlation over the quantization distortion. This covariance matrix is, however, a function of not only $\textbf{F}_\mathsf{BB}$, for which no estimate is available for the initial iteration, but also $\textbf{A}_\mathsf{B}$, which depends on $\textbf{F}_\mathsf{BB}$ too through \eqref{eq:A_Bussgang}. In order to facilitate the derivation, we therefore employ the AQNM-based covariance $\textbf{C}_{\textbf{q}_\mathsf{Q}\textbf{q}_\mathsf{Q}}$ given by \eqref{eq:Cqq_AQNM}, which is a direct function of $\textbf{F}_\mathsf{BB}$, while performing the power normalization in \eqref{eq:bb_normalization}  for the initial iteration. Unfortunately, obtaining $\textbf{C}_{\textbf{q}_\mathsf{Q}\textbf{q}_\mathsf{Q}}$ is still cumbersome as it should be computed along with $\textbf{F}_\mathsf{BB}$ because of the mutual dependency.

In \cite{Rupp2018EneEff}, the baseband precoder is derived assuming a sufficiently high-resolution DAC, and, hence, the covariance is simply assumed to be a zero matrix in power constraint computations. In our case, resolution of DAC is 1-bit, and the zero-covariance assumption (i.e., $\textbf{C}_{\textbf{q}_\mathsf{Q}\textbf{q}_\mathsf{Q}} \,{=}\, \textbf{0}_{N_\mathsf{RF}}$), hence, does not hold. We therefore propose a fixed-point iterative solution to find optimal  $\textbf{F}_\mathsf{BB}$ and $\textbf{C}_{\textbf{q}_\mathsf{Q}\textbf{q}_\mathsf{Q}}$ jointly \cite{Yapici2019QuanMassC}. As described in Algorithm~\ref{algorithm:overall}, we start 
by assuming $\textbf{C}_{\textbf{q}_\mathsf{Q}\textbf{q}_\mathsf{Q}} \,{=}\, \textbf{0}_{N_\mathsf{RF}}$, and recursively update $\textbf{F}_\mathsf{BB}$ by \eqref{eq:bb_normalization} and $\textbf{C}_{\textbf{q}_\mathsf{Q}\textbf{q}_\mathsf{Q}}$ by \eqref{eq:Cqq_AQNM} in sequence until a convergence condition (line \ref{algorithm:convergence_start} of Algorithm~\ref{algorithm:overall}) is achieved for the covariance matrix. In Section~\ref{sec:results}, we numerically verify a satisfactory convergence behavior for $\textbf{C}_{\textbf{q}_\mathsf{Q}\textbf{q}_\mathsf{Q}}$. 

\begin{algorithm}
  \caption{Alternating-Optimization Algorithm for Hybrid Precoder Design for 1-Bit DAC}
  \label{algorithm:overall}
    \begin{algorithmic}[1]
        \State \textbf{Initialize:} $\textbf{A}_\mathsf{Q}$ by \eqref{eq:A_AQNM}, $\textbf{C}_{\textbf{q}_\mathsf{Q}\textbf{q}_\mathsf{Q}}^{(\ell)} \gets \textbf{0}_{N_\mathsf{RF}}$, $\ell = {-}1,0$, $k \gets 1$, error tolerance $\epsilon$, number of iterations $K$
        
        \State \textbf{Initial Iteration}:
        \State Compute $\hat{\textbf{F}}_\mathsf{RF}$ by SVD of $\textbf{H}$
        \State Compute $\textbf{F}_\mathsf{RF}$ by applying APA~\cite{Heath2017HybArc} to $\hat{\textbf{F}}_\mathsf{RF}$
        \State Compute $\hat{\textbf{F}}_\mathsf{BB}$ by SVD of $\textbf{H}_\mathsf{e} \,{=}\, \textbf{H}\hspace{0.01in}\textbf{F}_\mathsf{RF} \textbf{A}_\mathsf{Q}$
        \While{$\| \textbf{C}_{\textbf{q}_\mathsf{Q}\textbf{q}_\mathsf{Q}}^{(k{-}1)} - \textbf{C}_{\textbf{q}_\mathsf{Q}\textbf{q}_\mathsf{Q}}^{(k{-}2)}
        \|_\mathsf{F} / N_\mathsf{RF} > \epsilon$} \label{algorithm:convergence_start}
            \State Update $\textbf{F}_\mathsf{BB}$ by \eqref{eq:bb_normalization} using $\textbf{C}_{\textbf{q}_\mathsf{Q}\textbf{q}_\mathsf{Q}}^{(k{-}1)}$ 
            \State Compute  $\textbf{C}_{\textbf{q}_\mathsf{Q}\textbf{q}_\mathsf{Q}}^{(k)}$ by \eqref{eq:Cqq_AQNM}
            \State $k \gets k+1$
        \EndWhile \label{algorithm:convergence_end}
         
        \State \textbf{Subsequent Iterations}:
        \For{$i = 2, \dots , K$}
            \State Update $\textbf{A}_\mathsf{B}$ and $\textbf{C}_{\textbf{q}_\mathsf{B}\textbf{q}_\mathsf{B}}$ by \eqref{eq:A_Bussgang} and \eqref{eq:Cqq_Bussgang} \label{algorithm:A_Cqq_Bussgang}
            \State Update $\hat{\textbf{F}}_\mathsf{BB}$ by SVD of $\textbf{H}_\mathsf{e} \,{=}\, \textbf{H}\hspace{0.01in}\textbf{F}_\mathsf{RF} \textbf{A}_\mathsf{B}$
            \State Update $\textbf{F}_\mathsf{BB}$ by \eqref{eq:bb_normalization} using $\textbf{C}_{\textbf{q}_\mathsf{B}\textbf{q}_\mathsf{B}}$
            \State Compute $\hat{\textbf{F}}_\mathsf{RF}$ by \eqref{eq:rf_optimization} \label{algorithm:rf_redesign_1}
            \State Compute $\textbf{F}_\mathsf{RF}$ by applying APA~\cite{Heath2017HybArc} to $\hat{\textbf{F}}_\mathsf{RF}$ \label{algorithm:rf_redesign_2}
        \EndFor
    \end{algorithmic}
\end{algorithm}

\vspace{-0.2in}

\section{Numerical Results} \label{sec:results}
In this section, we present the numerical results based on extensive Monte Carlo simulations to evaluate the performance of proposed hybrid precoder design assuming one-bit DAC. The simulation parameters are listed in Table \ref{table:simulation_parameters}. 
\begin{table}[!h]
\renewcommand{\arraystretch}{1.1}
\caption{Simulation Parameters}
\label{table:simulation_parameters}
\centering
\begin{tabular}{lc}
\hline
Parameter & Value \\
\hline\hline
Number of Tx antennas ($N_\mathsf{t}$) & 32 \\ 
Number of Rx antennas ($N_\mathsf{r}$) & 8 \\ 
Number of RF chains ($N_\mathsf{RF}$) & \{$4,8$\} \\ 
Number of streams ($N_\mathsf{s}$) & $N_\mathsf{RF}$ \\ 
Maximum transmit signal power ($\mathsf{P}_\mathsf{max}$) & $10$ Watts\\
Power of the unprecoded data ($P_\mathsf{s}$) & $1$ Watt\\
Number of clusters ($N_\mathsf{c}$) & $1$\\
Number of rays ($N_\mathsf{p}$) & $5$\\
Angular distribution of clusters & Uniform \\
Angular distribution of rays & Laplace\\
Angular spread of rays & $10^{\circ}$\\
Resolution of the phase shifters ($\Delta$) & $5^{\circ}$ \\
\hline
\end{tabular}
\end{table}
\begin{figure}[!t]
	\centering
	\vspace{-0.2in}
	\includegraphics[width=0.44\textwidth]{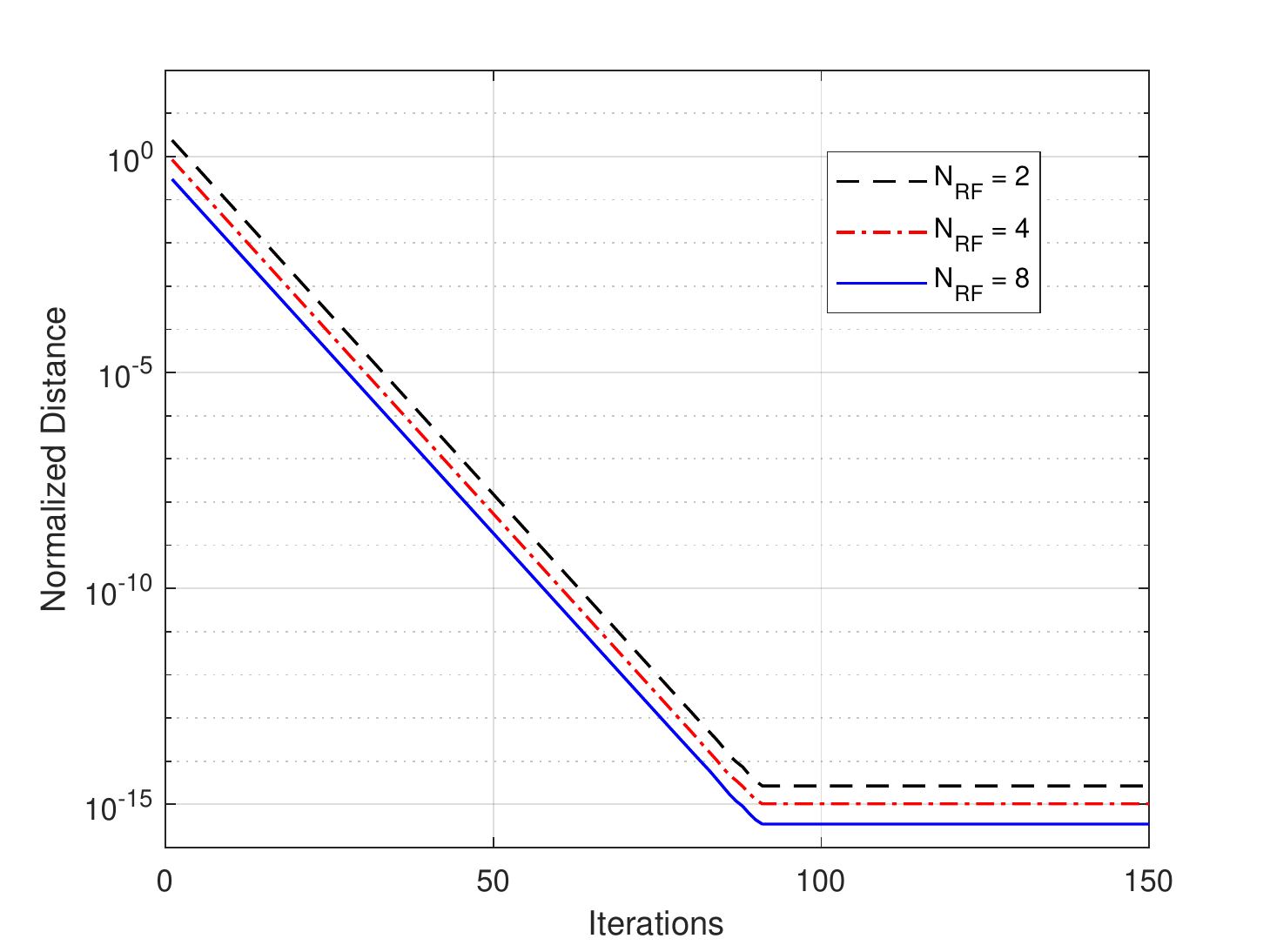}
	\vspace{-0.1in}
	\caption{Convergence of $\textbf{C}_{\textbf{q}_\mathsf{Q}\textbf{q}_\mathsf{Q}}$ for $N_\mathsf{RF} \,{\in}\, \{2,4,8\}$.}
	\label{fig:convergence}
	\vspace{-0.2in}
\end{figure}

In Fig.~\ref{fig:convergence}, we depict the convergence behavior of the covariance matrix $\textbf{C}_{\textbf{q}_\mathsf{Q}\textbf{q}_\mathsf{Q}}$ by lines \ref{algorithm:convergence_start}--\ref{algorithm:convergence_end} of Algorithm~\ref{algorithm:overall}. We observe that the normalized distance (i.e., line \ref{algorithm:convergence_start} of Algorithm~\ref{algorithm:overall}) converges (i.e., hits the precision limit of the computer in use) after a finite number of iterations for any choice of $N_\mathsf{RF}$. We therefore verify that it is possible to obtain a non-zero $\textbf{C}_{\textbf{q}_\mathsf{Q}\textbf{q}_\mathsf{Q}}$ through fixed-point iterations which produces more accurate rates since it does not ignore 
the quantization distortion.

\begin{figure}[!t]
\centering
\subfloat[$N_\mathsf{RF} \,{=}\, 4$]
{\includegraphics[width=0.45\textwidth]{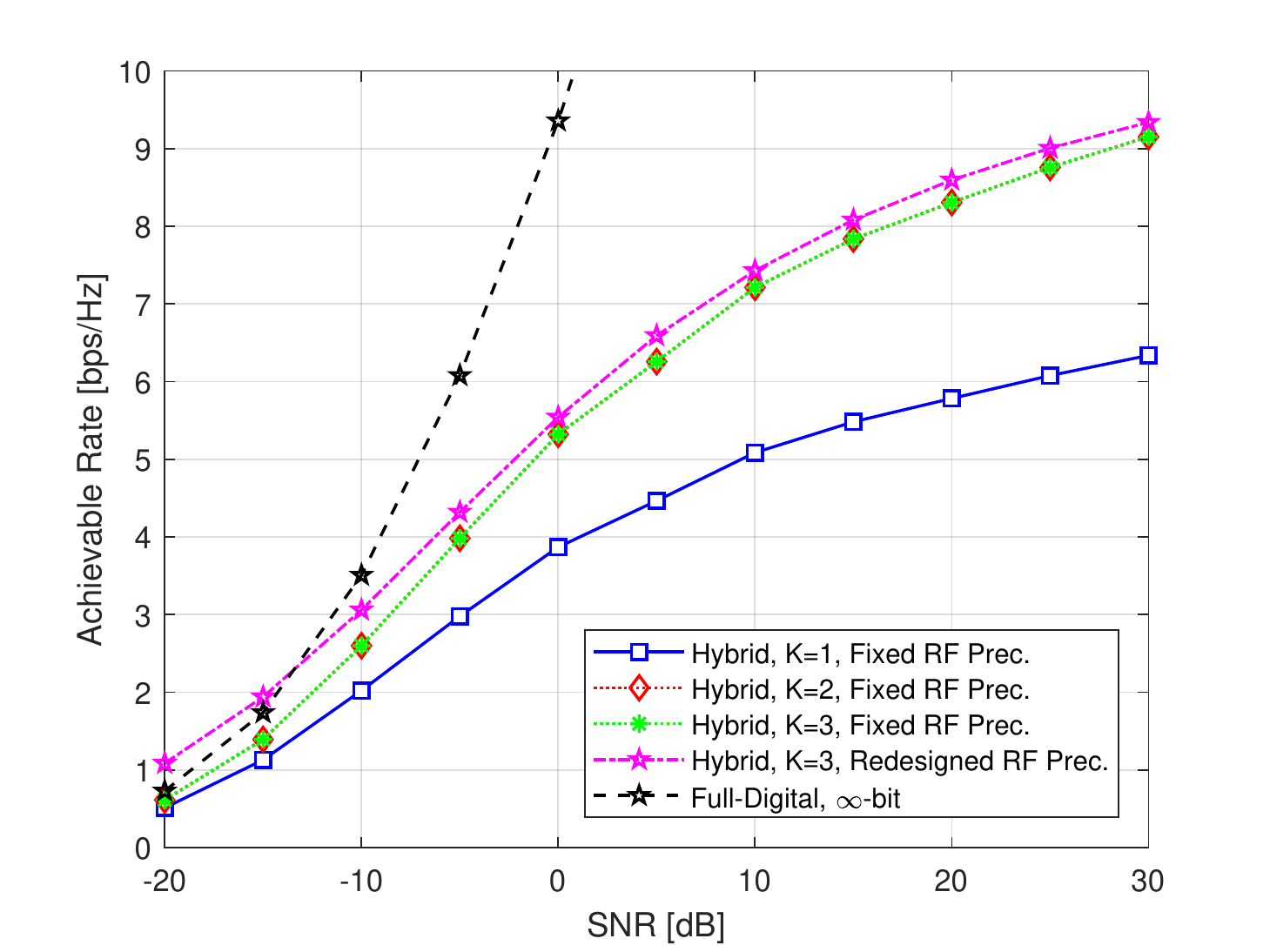}
\label{fig:Ns4}}\\
\vspace{-0.22in}
\subfloat[$N_\mathsf{RF} \,{=}\, 8$]
{\includegraphics[width=0.42\textwidth]{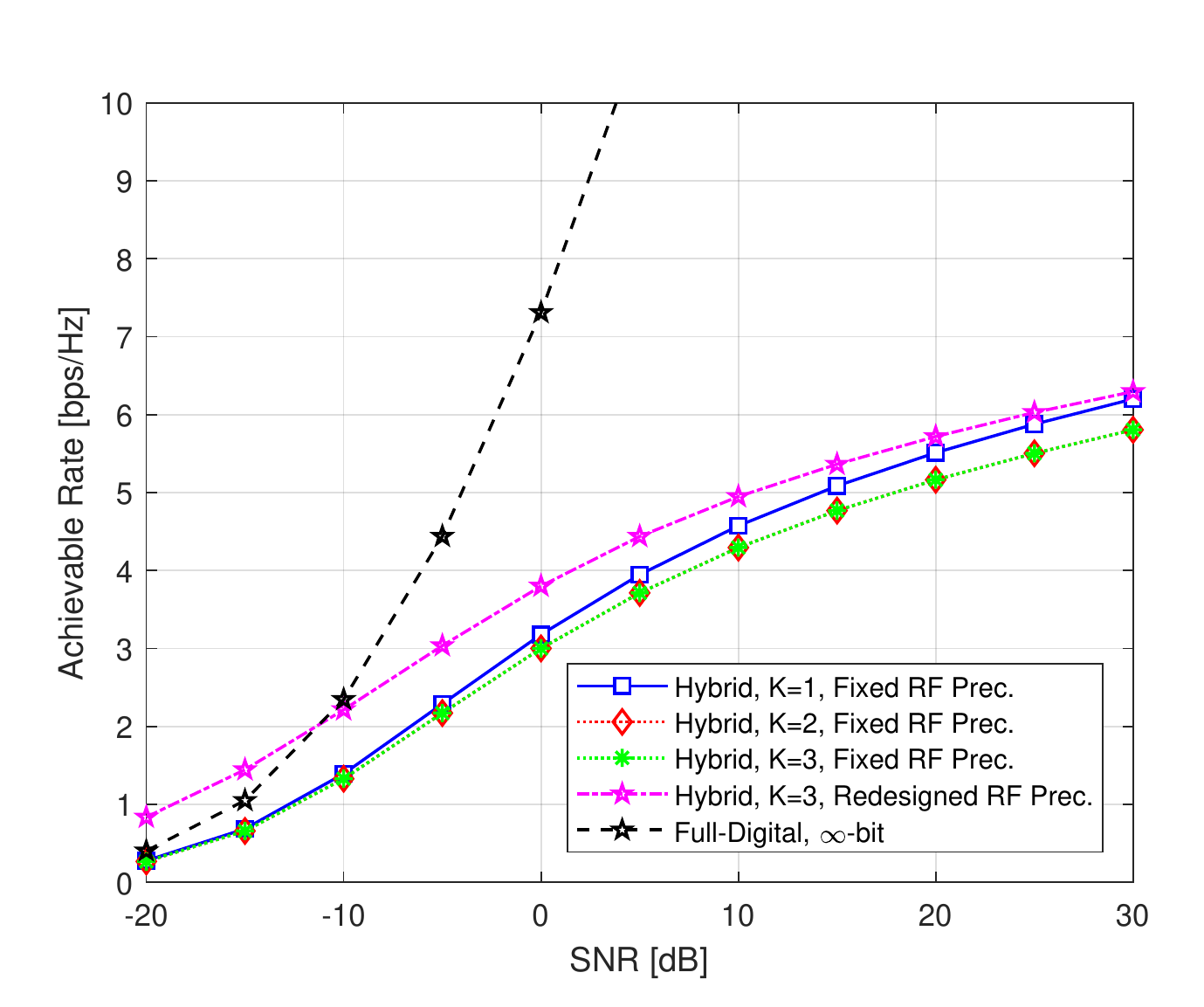}
\label{fig:Ns8}}
\vspace{-0.06in}
\caption{The achievable rate against varying SNR for $N_\mathsf{RF} \,{\in}\, \{4,8\}$, number of iterations $K \,{\in}\, \{1,2,3\}$, with fixed and redesigned RF precoder.}
\label{fig:rate}
\vspace{-0.3in}
\end{figure}

In Fig.~\ref{fig:rate}, we present the achievable rate results against varying signal-to-noise ratio (SNR), which is defined as $\mathsf{P}_\mathsf{max}/\sigma^2_\mathsf{n}$, for $N_\mathsf{RF} \,{\in}\, \{4,8\}$. For comparison purposes, we also include the performance of the SVD-based full-digital precoder $\textbf{F}_\mathsf{FD}$ with infinite-resolution DACs. Note that the performance is better in $N_\mathsf{RF} = 4$ than $N_\mathsf{RF} = 8$, since spatially correlated mmWave channel is not decomposed into full dimensional orthogonal sub-channels in the latter. 
In Fig.~\ref{fig:rate}\subref{fig:Ns4}, which assumes $N_\mathsf{RF} \,{=}\, 4$, the achievable rate improves after the initial iteration of the overall alternating-optimization algorithm, even when $\textbf{F}_\mathsf{RF}$ is kept the same (i.e., without implementing the lines \ref{algorithm:rf_redesign_1}-\ref{algorithm:rf_redesign_2} of Algorithm~\ref{algorithm:overall}). Since the initial iteration employs the AQNM-based weight and covariance matrices
, this performance improvement is basically due to the use of $\textbf{A}_\mathsf{B}$ and $\textbf{C}_{\textbf{q}_\mathsf{B}\textbf{q}_\mathsf{B}}$, which are both based on the Bussgang theorem (i.e., line \ref{algorithm:A_Cqq_Bussgang} of Algorithm~\ref{algorithm:overall}). We would like to remind that performance of even the initial iteration is associated with the recursive computation of the covariance $\textbf{C}_{\textbf{q}_\mathsf{Q}\textbf{q}_\mathsf{Q}}$ instead of assuming it is simply a zero matrix.

We also observe that redesigning $\textbf{F}_\mathsf{RF}$ (i.e., lines \ref{algorithm:rf_redesign_1}-\ref{algorithm:rf_redesign_2} of Algorithm~\ref{algorithm:overall}) improves the performance even further. Since $\textbf{F}_\mathsf{RF}$ redesign aims at directly maximizing the achievable rate by \eqref{eq:rf_optimization}, the respective performance is superior to that of the full-digital precoder at low SNR, which solely relies on SVD of the channel. We observe a similar behavior in Fig.~\ref{fig:rate}\subref{fig:Ns8} for $N_\mathsf{RF} \,{=}\, 8$. Since the number of multiplexed streams now doubles,
the subsequent iterations of the alternating-optimization algorithm without $\textbf{F}_\mathsf{RF}$ redesign cannot help improve the achievable rate, and even worsen it at high SNR. Despite that, $\textbf{F}_\mathsf{RF}$ redesign achieves even a larger performance gap than $N_\mathsf{RF} \,{=}\, 4$ in this challenging environment as compared to the fixed $\textbf{F}_\mathsf{RF}$ case.  

\vspace{-0.2in}
\section{Conclusion}
\label{sec:conclusion}

We propose a hybrid precoder for spatially correlated mmWave channels equipped with one-bit DAC and finite-quantized phase shifters. We carefully incorporate the impact of the quantization distortion, and obtain the baseband and RF precoders relying on the Bussgang theorem. The numerical results verify the superiority of the proposed alternating-maximization algorithm as compared to AQNM-based design.

\vspace{-0.15in}

\bibliographystyle{IEEEtran} 
\bibliography{IEEEabrv,bibfile}

\end{document}